\documentclass[]{aa}
\usepackage{graphicx}
\usepackage{setspace}
\usepackage{natbib}
\usepackage{longtable}
\usepackage{amssymb}
\usepackage[varg]{txfonts}
\bibpunct{(}{)}{;}{a}{}{,}

\begin{document}

\newcommand{\FeII}{[\ion{Fe}{ii}]}
\newcommand{\TiII}{[\ion{Ti}{ii}]}
\newcommand{\SII}{[\ion{S}{ii}]}
\newcommand{\OI}{[\ion{O}{i}]}
\newcommand{\OIp}{\ion{O}{i}}
\newcommand{\PII}{[\ion{P}{ii}]}
\newcommand{\NI}{[\ion{N}{i}]}
\newcommand{\NII}{[\ion{N}{ii}]}
\newcommand{\NIp}{\ion{N}{i}}
\newcommand{\NiII}{[\ion{Ni}{ii}]}
\newcommand{\CaIIp}{\ion{Ca}{ii}}
\newcommand{\PI}{[\ion{P}{i}]}
\newcommand{\CIp}{\ion{C}{i}}
\newcommand{\HeI}{\ion{He}{i}}
\newcommand{\MgIp}{\ion{Mg}{i}}
\newcommand{\MgIIp}{\ion{Mg}{ii}}
\newcommand{\NaI}{\ion{Na}{i}}
\newcommand{\HI}{\ion{H}{i}}
\newcommand{\brg}{Br$\gamma$}
\newcommand{\pab}{Pa$\beta$}

\newcommand{\macc}{$\dot{M}_{acc}$}
\newcommand{\lacc}{L$_{acc}$}
\newcommand{\lbol}{L$_{bol}$}
\newcommand{\mjet}{$\dot{M}_{jet}$}
\newcommand{\mh}{$\dot{M}_{H_2}$}
\newcommand{\Ne}{n$_e$}
\newcommand{\h}{H$_2$}
\newcommand{\kms}{km\,s$^{-1}$}
\newcommand{\um}{$\mu$m}
\newcommand{\lam}{$\lambda$}
\newcommand{\msyr}{M$_{\odot}$\,yr$^{-1}$}
\newcommand{\Av}{A$_V$}
\newcommand{\msun}{M$_{\odot}$}
\newcommand{\lsun}{L$_{\odot}$}
\newcommand{\cm}{cm$^{-3}$}
\newcommand{\ergscm}{erg\,s$^{-1}$\,cm$^{-2}$}

\newcommand{\bet}{$\beta$}
\newcommand{\alfa}{$\alpha$}

\hyphenation{mo-le-cu-lar pre-vious e-vi-den-ce di-ffe-rent pa-ra-me-ters ex-ten-ding a-vai-la-ble excited}

\title{The wind and the magnetospheric accretion onto the T~Tauri star S Coronae Australis at sub-au resolution}

\author{GRAVITY Collaboration(\thanks{GRAVITY is developed in a
collaboration by the Max Planck Institute for Extraterrestrial Physics,
LESIA of Paris Observatory and IPAG of Université Grenoble Alpes / CNRS,
the Max Planck Institute for Astronomy, the University of Cologne, the
Centro Multidisciplinar de Astrofisica Lisbon and Porto, and the European
Southern Observatory.}): R. Garcia Lopez \inst{1,2} \and K. Perraut \inst{3} \and A. Caratti o Garatti\inst{1,2} \and B. Lazareff\inst{3} \and J. Sanchez-Bermudez\inst{1} \and  M. Benisty\inst{3,13} \and C. Dougados\inst{3} \and L. Labadie\inst{4} \and W. Brandner\inst{1} \and P.J.V. Garcia\inst{5,6} \and Th. Henning\inst{1} \and T.P. Ray\inst{2} R.~Abuter \inst{7} \and A.~Amorim \inst{6} \and N.~Anugu \inst{6} \and J.P.~Berger \inst{7,3} \and
H.~Bonnet \inst{7} \and A.~Buron \inst{8} \and P. Caselli \inst{8} \and Y.~Cl\'enet \inst{9} \and V.~Coud\'e~du~Foresto \inst{9} \and W.~de~Wit \inst{10} \and C.~Deen \inst{8} \and F.~Delplancke-Ströbele \inst{7} \and J.~Dexter \inst{8} \and A.~Eckart \inst{4,11} \and F.~Eisenhauer \inst{8} \and C.E.~Garcia Dabo \inst{7} \and E.~Gendron \inst{9} \and
R.~Genzel \inst{8,12} \and S.~Gillessen \inst{8} \and X.~Haubois \inst{10} \and M.~Haug \inst{8,7} \and F.~Haussmann \inst{8} \and S.~Hippler \inst{1} \and Z.~Hubert \inst{1,9} \and C.A.~Hummel \inst{7} \and M.~Horrobin \inst{4} \and L.~Jocou \inst{3} \and
S.~Kellner \inst{8,11} \and P.~Kervella \inst{9,13} \and M.~Kulas \inst{1} \and
J.~Kolb \inst{10} \and S.~Lacour \inst{9} \and J.-B.~Le~Bouquin \inst{3} \and
P.~L\'ena \inst{9} \and M.~Lippa \inst{8} \and A.~M\'erand \inst{7} \and
E.~M\"uller \inst{1,10} \and T.~Ott \inst{8} \and J.~Panduro \inst{1} \and
T.~Paumard \inst{9} \and G.~Perrin \inst{9} \and O.~Pfuhl \inst{8} \and
A.~Ramirez \inst{7} \and C.~Rau\inst{8} \and R.-R.~Rohloff \inst{1} \and
G.~Rousset \inst{9} \and S.~Scheithauer \inst{1} \and M.~Schöller \inst{7} \and
C.~Straubmeier \inst{4} \and E.~Sturm \inst{8} \and W.F. Thi \inst{8} \and
E. van Dishoeck \inst{8,14} \and F.~Vincent \inst{9} \and I.~Waisberg \inst{8} \and
I.~Wank \inst{4} \and E.~Wieprecht \inst{8} \and M.~Wiest \inst{4} \and E.~Wiezorrek \inst{8} \and J.~Woillez \inst{7} \and S.~Yazici \inst{8,4} \and G.~Zins \inst{10}
}

   \institute{Max Planck Institute for Astronomy, K\"{o}nigstuhl 17, Heidelberg, Germany, D-69117
         \and
             Dublin Institute for Advanced Studies, 31 Fitzwilliam Place, D02\,XF86 Dublin, Ireland
             \and 
             Univ. Grenoble Alpes, CNRS, IPAG, F-38000 Grenoble, France
             \and
             I. Physikalisches Institut, Universität zu Köln, Zülpicher Str. 77, 50937, Köln, Germany
             \and
Universidade do Porto - Faculdade de Engenharia, Rua Dr. Roberto Frias, 4200-465 Porto, Portugal
\and
CENTRA, Instituto Superior Tecnico, Av. Rovisco Pais, 1049-001 Lisboa, Portugal
\and European Southern Observatory, Karl-Schwarzschild-Str. 2, 85748
Garching, Germany
\and
Max Planck Institute for Extraterrestrial Physics, Giessenbachstrasse, 85741 Garching bei M\"{u}nchen, Germany
\and
LESIA, Observatoire de Paris, PSL Research University, CNRS, Sorbonne Universit\'es, UPMC Univ. Paris 06, Univ. Paris Diderot, Sorbonne Paris Cit\'e, France
\and
European Southern Observatory, Casilla 19001, Santiago 19, Chile
\and
Max-Planck-Institute for Radio Astronomy, Auf dem H\"ugel 69,
53121 Bonn, Germany
\and 
Department of Physics, Le Conte Hall, University of California, Berkeley, CA 94720, USA
\and 
Unidad Mixta Internacional Franco-Chilena de Astronomía (CNRS UMI 3386), Departamento de Astronomía, Universidad de Chile, Camino El Observatorio 1515, Las Condes, Santiago, Chile
\and 
Sterrewacht Leiden, Leiden University, Postbus 9513, 2300 RA Leiden, The Netherlands\\
             \email{rgarcia@cp.dias.ie; karine.perraut@univ-grenoble-alpes.fr}
              }

   \date{Received ; accepted }

\titlerunning{Tracing the wind and the magnetospheric accretion of S\,CrA at sub-au resolution}
\authorrunning{Garcia Lopez, R. et al.}

  \abstract
   {}
   {To investigate the inner regions of protoplanetary disks, we performed near-infrared interferometric observations of the classical T~Tauri binary system S\,CrA.}
   {We present the first VLTI-GRAVITY high spectral resolution ($R$~$\sim$~4000) observations of a classical T~Tauri binary, S\,CrA (composed of S\,CrA\,N and S\,CrA\,S and separated by $\sim$1\farcs4), combining the four 8-m telescopes in dual-field mode. }
   {Our observations in the near-infrared K-band continuum reveal a disk around each binary component, with similar half-flux radii of about 0.1 au at d$\sim$130\,pc, inclinations ($i=$28$\pm$3\degr\ and $i=$22$\pm$6\degr), and position angles (PA=0\degr$\pm$6\degr\ and PA=-2\degr$\pm$12\degr), suggesting that they formed from the fragmentation of a common disk. The S\,CrA\,N spectrum shows bright \HeI\ and \brg\ line emission exhibiting inverse P-Cygni profiles, typically associated with infalling gas. The continuum-compensated \brg\ line visibilities of S\,CrA\,N show the presence of a compact \brg\ emitting region the radius of which is about $\sim$0.06\,au, which is twice as big as the truncation radius. This component is mostly tracing a wind. Moreover, a slight radius change between the blue- and red-shifted \brg\ line components is marginally detected.} 
   {The presence of an inverse P-Cygni profile in the \HeI\ and \brg\ lines, along with the tentative detection of a slightly larger size of the blue-shifted \brg\ line component, hint at the simultaneous presence of a wind and magnetospheric accretion in S\,CrA\,N.
}


   \maketitle
%
%-------------------------------------------------------------------

\section{Introduction}

The physical structure and processes acting in the inner regions of protoplanetary disks are still poorly constrained, yet they are key elements for understanding global disk dynamics and evolution. In this region, accretion flows, winds and outflows are essential to control angular momentum, alter the gas content and drive the dynamics of the gas \citep[see e.g. ][]{alexander14}. While spectroscopic studies can bring global constraints to these processes, only spatially resolved observations with milliarcsecond (mas) resolution, that is, on sub-astronomical unit (au) scale for the nearest star forming regions, can discriminate competing models. 

In the past decade, near-infrared (NIR) interferometers have successfully resolved the inner regions of circumstellar disks around young stars. Detailed studies of individual objects were done through observations of the dust, traced by the continuum emission \citep{tannirkulam08a,benisty10,kraus13}, or of the hot and warm gas components through hydrogen recombination and CO vibrational lines \citep{tatulli07,eisner_herbig07,kraus08}. The first statistical analysis of the dust was carried out recently with the advent of the 4-telescope H-band instrument PIONIER at the Very Large Telescope Interferometer (VLTI). This survey showed that the sublimation front is in general rather smooth and wide, in agreement with the presence of multiple grain populations, and that its thickness is larger than predicted by hydrostatic models \citep{PIONIER_TTauri,lazareff17}. 

However, observations of the gaseous disk component are more challenging and they have so far been restricted to a handful of bright and/or highly accreting sources \citep[e.g.][]{kraus08, eisner14, rebeca15, ale15}. In this context, the \HI\,\brg\ line has been the focus of most of the interferometric studies performed in the $K$ band so far to trace the hot inner gas, and thus, the physical processes taking place within the inner disk region. These studies, mainly focused on Herbig stars, have shown that, in most cases, the \brg\ line originates from a very compact region located well within the dust sublimation radius. Despite its small angular size (a few tenths of mas), most of the detailed individual studies associate the origin of this line with a wind or the hot layers of a circumstellar disk rather than with magnetospheric accretion \citep{weigelt11,ellerbroek14,mendigutia15,rebeca15, ale15,rebeca16}. Results from these highly irradiated disks are, however, probably not representative of the bulk of the young stellar object (YSO) population, and in particular of the Classical T Tauri stars (CTTSs), which have much lower stellar temperatures than their higher mass counterparts, the Herbig Ae/Be stars. By modelling spatially and spectrally ($R$~$\sim$~2000) resolved \brg\ line profiles of CTTSs, \citet{eisner10} suggest that the \HI\, gas mostly traces infalling columns of material, although some emission might also originate from compact regions of outflowing winds.

In this paper, we present the first interferometric observations of a CTTS obtained in the K band with GRAVITY, the second generation beam combiner at the VLTI \citep{Eisenhauer2011,GRAVITY}, and a spectral resolution of $R$~$\sim$~4000. We observed the binary system S\,CrA located at a distance of $\sim$130\,pc \citep{prato03, neuhauser08}, with a separation of $\sim$1\farcs4 and a PA of 157\degr\ \citep{reipurth93_binaries,prato03}. Both sources are CTTSs and they appear coeval \citep{prato03,petrov14_SCra}. Due to the high veiling measured in both objects, the spectral types are uncertain ranging from G to M \citep{prato03, carmona07, petrov14_SCra}. Both stars are strong accretors. They possess flared disks and a remnant envelope, as suggested by Herschel observations \citep{sicilia-aguilar13}. The system is still very active and associated with HH\,82 (PA$\sim$95$\degr$) and HH\,729 (PA$\sim$115$\degr$) \citep[see][]{wang,kumar,peterson11}.

S\,CrA was early classified as one of the brightest examples of the YY\,Ori objects, a class of pre-main sequence stars, first identified by \citet{walker}. These objects show a broad red-shifted absorption component (inverse P Cygni profile) on the wings of selected emission lines, indicating infalling gas at velocities larger than 100\,km\,s$^{-1}$. 
Several detailed spectroscopic studies of the optical \HI\ lines in S\,CrA are reported in literature~\citep[see e.g.][]{bertout82,appenzeller86,krautter90}. In particular, \citet{krautter90} favour the model of magnetospheric accretion for the origin of these red-shifted absorption features.

We first describe the GRAVITY observations and data processing (Sect.\,\ref{sect:observations}), then the spectroscopic and interferometric results (Sect.\,\ref{sec:results}). We interpret them using geometric models applied to the continuum and the \brg\ line (Sect.\,\ref{sec:models}). Finally, all findings are discussed in Sect.\,\ref{sect:discussion}.

\section{Observations and data reduction}
\label{sect:observations}

%%%%%%%%%%%%%%%%%%%%%%%%%%%%%%%%%%%%%%%%%%%%%%%%%%%%%%%%%%%%%%%%%%%%%%%%%%%%%%%%%%%%%%%%%%%%%%%%%%%%%%%%%%%%%%%%
\begin{table*}[t]
\tiny
\caption{\label{tab:obslog} Observation log of the VLTI/GRAVITY high-resolution ($R$~$\sim$4000) observations of S\,CrA North and South conducted with the UTs.}
\centering
\vspace{0.1cm}
\begin{tabular}{@{}c c c c c c c | c c }
\hline \hline
UT Date & Tot. Int. S\,CrA N &  Tot. Int. S\,CrA S & DIT\tablefootmark{a} & NDIT\tablefootmark{b} & 
Proj. baselines & PA\tablefootmark{c} & Calibrator & UD diameter\tablefootmark{d}\\ 
 & [s] & [s] & [s] & \# &  [m] & [$\circ$] & &[mas]\\  \hline 
2016-07-20 & 2100 & 600 & 60 & 5  &  42, 52, 60, 81, 93, 115 & 53, 38, 130, 100, 45, 76  & HD\,188787 & $0.252 \pm 0.018$\\ 
2016-08-16 &  1200 & 1200 & 60 & 5  &  47, 56, 59, 87, 102, 130 & 31, 18, 100, 70, 24, 50  & HD\,176047 & $0.309 \pm 0.021$\\
\hline
\end{tabular}
\tablefoot{
\tablefoottext{a}{Detector integration time per interferogram.}
\tablefoottext{b}{Number of interferograms.}
\tablefoottext{c}{Baseline position angle from the shortest to longest baseline.}
\tablefoottext{d}{The calibrator uniform-disk (UD) diameter (\textit{K} band) was taken from \cite{lafrasse10}
.}
}
\end{table*}
%%%%%%%%%%%%%%%%%%%%%%%%%%%%%%%%%%%%%%%%%%%%%%%%%%%%%%%%%%%%%%%%%%%%%%%%%%%%%%%%%%%%%%%%%%%%%%%%%%%%%%%%%%%%%%%%%%%%%%%%%%%%%%

The binary system S\,CrA was observed in two runs (20th of July and 16th of August 2016, see Table\,\ref{tab:obslog}) with the VLTI instrument GRAVITY, using the four 8-m Unit Telescopes (UTs) of the European Southern Observatory and recording interferograms on 6 baselines simultaneously. The observations were performed in dual-field mode swapping between the North (S\,CrA\,N) and South (S\,CrA\,S) components of the binary system. This allowed us to alternate each of the binary components on the fringe tracker (FT) and science (SC) detectors. 

The visibilities and closure phases for the FT detector are recorded at low spectral resolving power (${\rm R}$~$\sim$~23, e.g., 5 spectral channels over the K band) with the FT working at 1~kHz and allowing the atmospheric effects on the fringe patterns to be frozen. On the other hand, the SC detector was set at high spectral resolution (HR; ${\rm R}$~$\sim$~4000, i.e., a radial velocity (${\rm RV}$) resolution of $\Delta {\rm RV} \sim$70\,\kms) in the combined polarization mode. More details about the GRAVITY instrument and observing modes can be found in \cite{GRAVITY}. A detailed log of the observations is reported in Table\,\ref{tab:obslog}.

%%%%%%%%%%%%%%%%%%%
\begin{figure*}[t]
	\centering
	\includegraphics[width=16 cm]{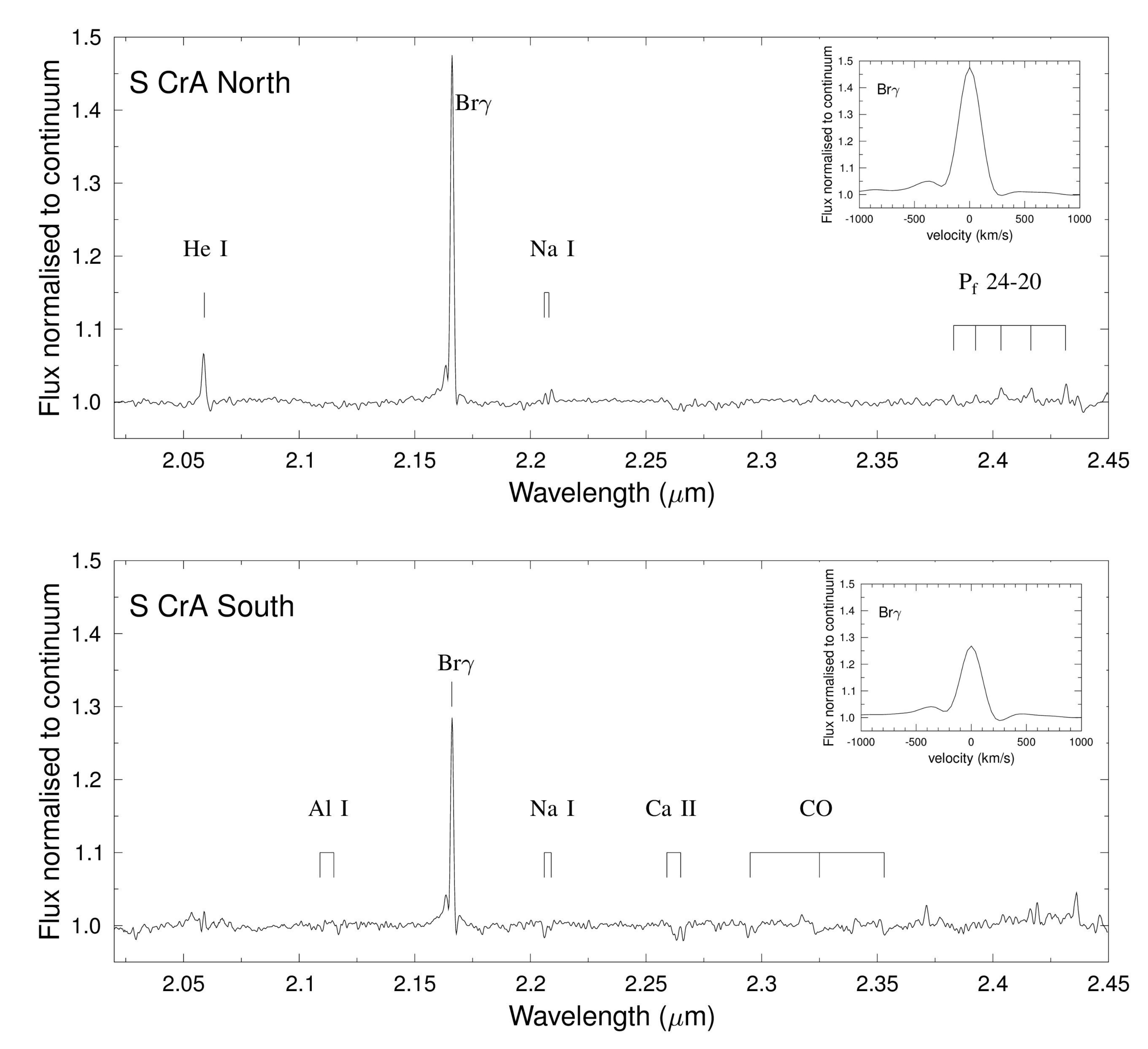}
    \caption{GRAVITY observations of S\,CrA taken during the July campaign showing the spectra of S\,CrA\,North (top) and South (bottom). Insets display the velocity calibrated \brg\ line profiles.}
    \label{fig:spec}
\end{figure*}
%%%%%%%%%%%%%%%%%%%%%%%

The data were reduced using the GRAVITY Data Reduction Software v0.9 \citep{DRS}. To calibrate the atmospheric transfer function the calibrators HD\,188787 and HD\,176047 were also observed with GRAVITY in single-field mode (Table\,\ref{tab:obslog}) at the same airmass as the scientific target. To obtain the spectra of S\,CrA\,N and S\,CrA\,S, we performed an average of the HR spectrum recorded in each of the four photometric channels. The calibrators were also used as telluric standards after removing the intrinsic photospheric absorption features present in their spectra and dividing them by a black-body at the appropriate temperature. The telluric corrected spectra of S\,CrA\,N and S\,CrA\,S were then flux calibrated assuming K-band magnitudes of 6.6 and 7.3, respectively \citep{prato03}. The wavelength calibration of the GRAVITY spectra was refined using the many telluric features present in the whole K-band, obtaining an average shift of $\sim$3\AA. The residual uncertainty is about 1--1.5\AA\, or $\sim$14--20\,km\,s$^{-1}$.

The weather conditions in the August run were not optimal and thus the signal-to-noise ratio of the HR science observations was much worse than those of the July run. Therefore while we can use all the FT low-resolution datasets to fit the continuum of each binary system component (Sect.\,\ref{sub:model_continuum}), we can only use the July HR observations of S\,CrA\,N to model the \brg\ line emitting region of this component (Sect.\,\ref{sub:model_BrG}). The spectro-interferometric data for S\,CrA\,S around the \brg\ line have a poor signal-to-noise ratio even in the July run and therefore they are only displayed in the Appendix and could not be examined any further. Finally, even with the good-quality July dataset of S\,CrA\,N, we were not able to retrieve differential interferometric signals across the \HeI\ line (Sect.~\ref{sec:spec}), because of its small line-to-continuum ratio ($<$10\%).

\section{Results}
\label{sec:results}
VLTI-GRAVITY observations allow us to obtain the K-band spectra (from $\sim$2.0\,\um\ to $\sim$2.5\,\um), as well as six spectrally dispersed visibility amplitudes 
and differential phases, and three closure phase measurements. 

\begin{figure*}[t]
	\centering
	\includegraphics[width=9.1 cm]{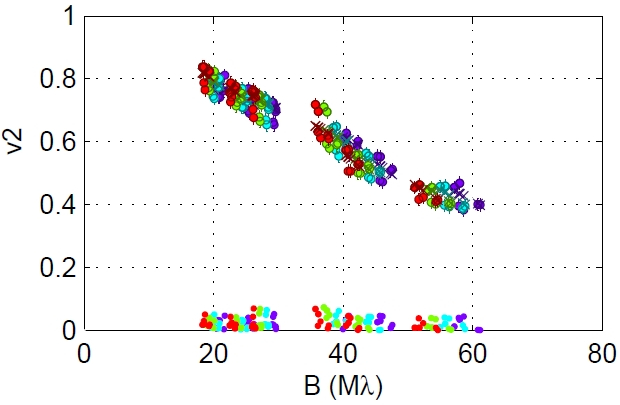}
    \includegraphics[width=9.1 cm]{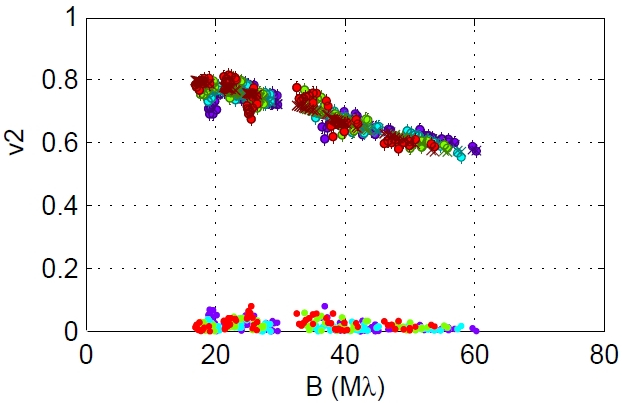}
    \caption{
    Squared visibility ($V^2$) vs. projected baseline in units of 10$^6\lambda$ for four FT spectral channels from our GRAVITY+UT observations of S\,CrA North (\emph{left panel}) and South (\emph{right panel}). The data and best fitting model are represented by filled symbols and crosses, respectively. The absolute value of the fit residuals is reported at the bottom of each panel. Different colours indicate different spectral channels.}
    \label{fig:continuum_visibilities}
\end{figure*}

\subsection{Spectrum of each component}
\label{sec:spec}

The full K-band spectrum of S\,CrA\,N is shown in Fig.\,\ref{fig:spec}-top. It displays bright \HeI\, 2.06\,\um\ and \brg\ emission, as well as a faint \NaI\, doublet (at 2.206\,$\mu$m and 2.209\,$\mu$m) and \HI\,Pfund (24-20) emission lines. These lines are usually found in the spectra of actively accreting CTTSs. Interestingly, the line profile of the \HeI\ line and, marginally, that of the \brg\ line show an inverse P-Cygni line profile with the red-shifted absorption component at radial velocities of 400$\pm$50\,\kms\ and 300$\pm$100\,km\,s$^{-1}$, respectively. This feature is usually associated with infalling gas located in magnetospheric accretion columns \citep{edwards06, kurosawa11}.
Notably, the red-shifted absorption features of the  \HI\ Balmer lines were detected at 220--270\,km\,s$^{-1}$, extending out to 350--400\,km\,s$^{-1}$~\citep[see][]{bertout82,krautter90}.

Moreover, the \brg\ line displays a double peaked asymmetric profile with radial velocities of $\sim$0\,\kms\ and $\sim$-360$\pm$40\,\kms, where the blue-shifted component is likely tracing a high-velocity jet or wind. 

The K-band spectrum of S\,CrA\,S (Fig.\,\ref{fig:spec}-bottom) also exhibits a double peaked and inverse P-Cygni \brg\ line and several photospheric absorption lines (Al I 2.110-2.117), NaI (2.207-2.209), CaI (2.226-2.227) and CO bandheads (2.29 - 2.4), indicative of a late type (K late or early M) star.

\subsection{Continuum visibilities of both components}
\label{sub:continuum}

The dual-field mode swapping between the North and South components of the binary system allows us to measure and compare the continuum visibilities recorded with both FT and SC detectors. Good agreement at a level better than 5\% was found between the calibrated continuum visibilities measured on the FT and SC detectors in July. Hereafter we use the FT data sets to fit the continuum of both components (Sect.\,\ref{sub:model_continuum}). The FT square visibility values are shown in Fig.\,\ref{fig:continuum_visibilities}. 

\subsection{Spectro-differential visibilities, phases, and closure phases of S\,CrA\,N}

\begin{figure*}[t]
	\centering
	\includegraphics[width=17 cm]{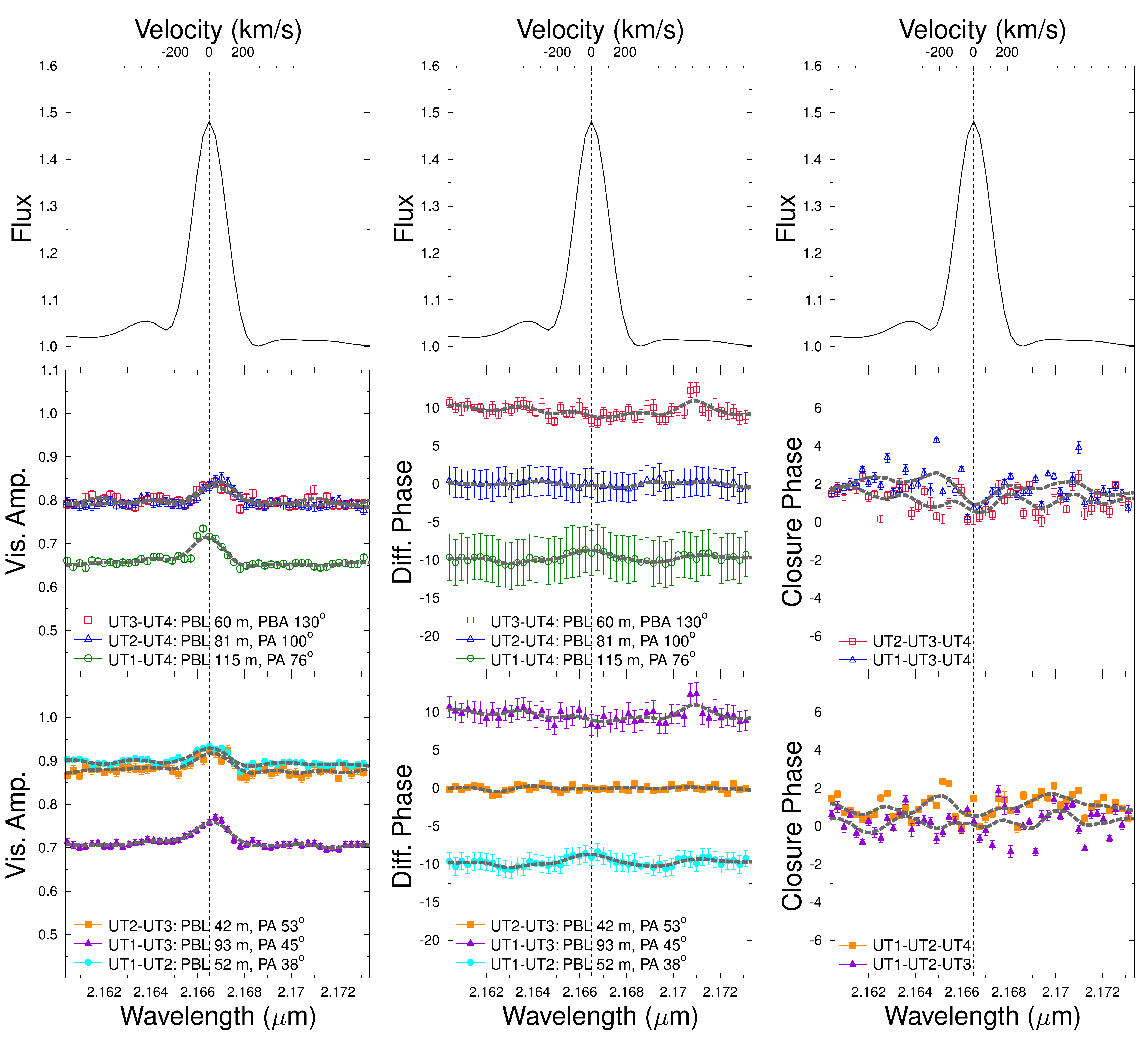}
    \caption{GRAVITY+UT HR observations of S\,CrA\,N on 2016-07-20. The \brg\ line profile normalised to the continuum is represented on top of each panel. The radial velocity is with respect to the LSR. Dashed lines represent smoothed data by two spectral channels. {\bf Left:} Wavelength-dispersed
    visibility amplitudes measured in the UT3-UT4, UT2-UT4, and UT1-UT4 (middle panel), and the UT2-UT3, UT1-UT3, and UT1-UT2 (bottom) baselines. {\bf Middle:} Same as ${\rm left}$ but for the wavelength-dispersed differential-phase signals. For clarity, the differential phases of the first and last baselines have been shifted by +10$^\circ$ and -10$^\circ$. {\bf Right:}  Wavelength-dispersed closure phase signals
    for the triplets UT2-UT3-UT4 and UT1-UT3-UT4 (middle panel), and UT1-UT2-UT4 and UT1-UT2-UT3 
    (bottom panel). }
    \label{fig:raw_data}
\end{figure*}

Figure\,\ref{fig:raw_data} displays all the interferometric observables around the \brg\ line. The spectrally dispersed visibilities increase in the \brg\ line relative to the continuum for all the baselines, indicating that the \brg\ emission originates in a region more compact than the continuum emitting region. Similar results were found in previous spectro-interferometric studies of Herbig Ae/Be stars, as well as classical T~Tauri sources \citep[see e.g.][]{kraus08, eisner14, ale15, rebeca15, kurosawa16}.

Interestingly, our results show a slight displacement of the peak of the total visibilities within the \brg\ line with respect to the peak of the line (see the dotted vertical line in Fig.\,\ref{fig:raw_data}), as well as a change in the shape of the visibilities from baseline to baseline. This suggests the presence of different physical structures in the environment of S\,CrA\,N with different kinematic signatures. It should be, however, mentioned that the shift in the peaks of the visibilities within the \brg\ line is of the order of a few tens of \kms, that is, the shifts are smaller than the spectral resolution of our GRAVITY observations. 

Mean closure phases of $\sim$2$^\circ$ are present in the largest telescope triplets (UT1-UT3-UT4, UT2-UT3-UT4 and UT1-UT2-UT4), hinting to a slightly asymmetric continuum. This observation agrees well with the absolute closure phase signal of 2.73$^\circ\pm$2.5$^\circ$ previously detected in the PIONIER H-band observations of S\,CrA\,N \citep{PIONIER_TTauri}. Moreover, this closure phase signal decreases across the \brg\ line for the UT1-UT3-UT4 triplet. Even if the dispersion of the signal (of about 1$^\circ$ peak-to-valley) does not allow further conclusions, this might indicate a first detection of an asymmetry in the innermost regions of a CTTS.

\subsection{Continuum-compensated \brg\ line visibilities of S\,CrA\,N}
\label{sub:pure_line_visibilities}

Continuum-compensated (or pure line) visibilities can be derived by considering that the visibilities across the \brg\ line have contributions from both the line emitting component and the continuum component. In this case, following the method described in \cite{weigelt07}, the \brg\ line continuum-compensated visibility can be derived from:
\begin{equation}
 V_{Br\gamma} = \frac{\sqrt{\textbar F_{tot} V_{tot} \textbar^2 + \textbar F_{cont} V_{cont} \textbar^2 - 2 F_{tot} V_{tot}F_{cont} V_{cont} cos \phi} }{F_{line}}
\end{equation}
where $F_{tot} = F_{cont} + F_{line}$, $F_{line}$, and $F_{cont}$ are the total, the line, and the continuum fluxes. $V_{cont}$ and $V_{tot}$ are the estimated continuum visibility and the observed total visibility, respectively, and $\phi$ is the differential phase measured within the line.

\begin{figure}
	\includegraphics[width=9cm]{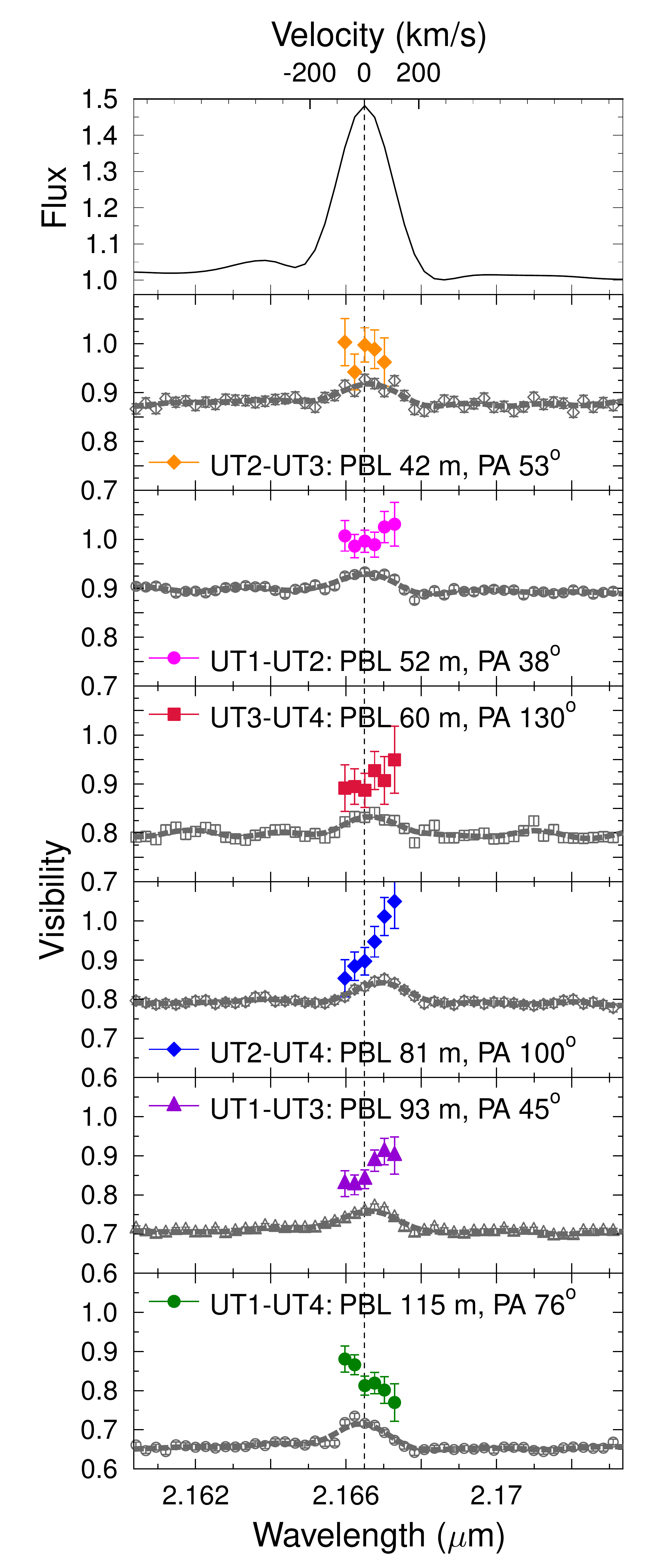}
    \caption{S\,CrA\,N continuum-compensated pure \brg\ line visibilities (full coloured symbols) computed for each of the six GRAVITY baselines. As a reference, the \brg\ line profile normalised to the continuum is displayed in the top panel. The wavelength dispersed visibility amplitudes are overplotted for each PBL (empty grey symbols), 
    along with a spectral binning of the visibility amplitudes (solid grey line). Velocities are with respect to the LSR.}
    \label{fig:pure_line_visibilities_all_channels}
\end{figure}

Our GRAVITY HR observations allow us to measure the pure \brg\ line visibilities for six spectral channels, namely those with a line-to-continuum ratio larger than 10\% (see Fig.\,\ref{fig:pure_line_visibilities_all_channels}). Therefore our analysis includes the main spectral component, but, unfortunately, excludes the high velocity components. The continuum-compensated \brg\ visibilities reveal a change in behaviour from the shortest to the longest baselines. For the three shortest baselines, both blue- and red-shifted line components are unresolved (V$\sim$1). The three longest baselines resolve at least one of the two components (V$<$1), with the blue-shifted component being more extended than the red-shifted one for the 81\,m and 93\,m projected baselines, whereas the opposite is observed for the 115\,m projected baseline.

\section{Modelling}
\label{sec:models}

\subsection{K-band continuum of S\,CrA\,N and S}
\label{sub:model_continuum}

\begin{table*}[t]
\caption{Parameters derived from the best fit of the FT continuum data for both S\,CrA components (see Fig.~\ref{fig:continuum_visibilities}). See text for parameter definition. }
\label{tab:fitCont}
\centering
\begin{tabular}{c c c c c c c c c c c c }
\hline
{\bf SCrA} & Sp. & $T_{\rm eff}$ & $k_s$ & $k_c$ & $f_s$ & $f_h$ & $f_c$ & $i$  & $PA$\tablefootmark{a} & $a$ & $a$ \\ 
& Type & [K] & & & & & & [$^\circ$] & [$^\circ$] & [mas] & [au]  \\
\hline\\
North & G0 & 5920 & 1.35 & -3.73$\pm$1.27 & 0.30$\pm$0.04 & 0.06$\pm$0.01 & 0.64$\pm$0.04 & 28$\pm$3 & 0$\pm$6 & 0.83$\pm$0.04 & 0.108$\pm$0.005 \\
South & K0 & 5280 & 1.25 & -4.03$\pm$0.74 & 0.52$\pm$0.02 & 0.10$\pm$0.01 & 0.38$\pm$0.02 & 22$\pm$6 & -2$\pm$12 & 0.70$\pm$0.04 & 0.090$\pm$0.005 \\
\hline
\end{tabular}
\tablefoot{The reduced $\chi^2$ values of the best fit model are 1.71 and 1.35 for the North and South component, respectively. 
\tablefoottext{a}{The PA corresponds to the major axis of the disk with North up and East to the left.}
}
\end{table*}

We used the modelling tool developed for the PIONIER instrument to fit the FT data on both components of the binary system \citep[for further modelling details refer to][]{lazareff17}. 
The effective temperature and photospheric index $k_s$ at the central wavelength of the K-band for each of the S\,CrA components were taken from \cite{pecaut13}, assuming spectral types of G0 and K0 for the North and South components, respectively. 

The visibility values at the shortest baseline, especially for the South component (see Fig.\,\ref{fig:continuum_visibilities}), can not be described by a pure Gaussian radial profile, but is described better by e.g. a Gaussian plus a halo component or a Lorentzian profile. In this case, we have focused our analysis on Gaussian plus halo fitting to be consistent with the \brg\ analysis (see Sect.\,\ref{sub:model_BrG}).
The continuum visibilities were then fitted by assuming a linear combination of the stellar component (unresolved at all the baselines), a halo (fully resolved at all baselines), and a circumstellar component, with the flux fraction of each component ($f_s$, $f_h$, and $f_c$ respectively) as the combination coefficients. 
The free fitting parameters are therefore two flux fractions among $f_{\rm s}$, $f_{\rm h}$, $f_{\rm c}$ (since the sum of the three fractions equals 1), the spectral index of the circumstellar environment $k_{\rm c}$ over the K band, the inclination $i$, the projected angle $PA$ (with respect to the major axis of the disk, with North up and East to the left), and $a$, the angular size (HWHM) of the circumstellar environment for a Gaussian radial profile. 

The results of the best fit for each S\,CrA component are summarised in Table~\ref{tab:fitCont}, and the best fit model for each source is overplotted in Fig.\,\ref{fig:continuum_visibilities}. Both components present the same inclinations (very close to face-on) and the same PAs (along the North-South direction). Considering a distance of 130~pc as \citet{prato03}, the angular sizes of the best fit model translate into 0.108~$\pm$~0.005~au and 0.090~$\pm$~0.005~au, which can be compared to the sublimation radius. According to the authors and the considered spectral type, the total luminosity of the S\,CrA binary system ranges between 3.0~L$_{\odot}$ \citep{prato03} and 5.5~L$_{\odot}$ \citep{Wilking1992}, which can be translated in a range of inner rim radius of 0.11-0.15~au for S\,CrA\,N and of 0.06-0.08~au for S\,CrA\,S. 
Our modelling leads to a fraction of total flux over stellar flux of 3.3$\pm$0.4 for the North component, and 1.92$\pm$0.07 for the South one. These values are consistent with the ones reported in \cite{prato03}, namely 2.89$\pm$1.60 and 3.15$\pm$2.00, for the North and South component, respectively.
Finally, we converted the spectral slope of the circumstellar component $k_c$ derived from interferometric data into a dust temperature: we obtained dust temperatures varying around 810-1200~K for S\,CrA\,N, and around 850-1050~K for S\,CrA\,S. These temperatures are lower than the \textit{a priori} dust sublimation temperature of 1500~K generally considered for the silicate, which is not surprising since a part of the observed circumstellar environment is beyond the dust sublimation radius.

\subsection{\brg\ line emitting region of S\,CrA\,N}
\label{sub:model_BrG}

To further investigate the apparent different dimensions of the red- and blue-shifted components of the \brg\ line, we performed a channel by channel geometric modelling of the pure \brg\ line visibilities shown in Fig.\,\ref{fig:pure_line_visibilities_all_channels}, using a Gaussian. 
First, we fitted the line visibilities with only the Gaussian semi-major axis $a$ as the only one free parameter, and assuming the same disk inclination and PAs as derived for the continuum (Table\,\ref{tab:fitCont}). 

Table\,\ref{tab:pure_line_visibilities} shows the size of the semimajor axis ($a_{Br\gamma}$) of our best Gaussian fittings for each spectral channel, along with the $\chi^2$ values of the fitting. Open squares in Fig.\,\ref{fig:fits_pure_line_visibilities}.2 show the results of our fitting for each baseline and spectral channel. 
In principle, the Br$\gamma$ line emission is not expected to have the same spatial distribution as the continuum, and indeed, the single-parameter fit shows that at least the angular size is different. Accordingly, similar series of fits were performed with three free parameters: $a$, $i$ and $PA$. The results are listed in Table\,\ref{tab:pure_line_visibilities} and shown in Fig.\,\ref{fig:fits_pure_line_visibilities}.2 (triangles). The figure shows that the three parameter fit is only slightly better than the one parameter fit. However, its reduced $\chi^2_r$ is worse because of the larger number of free parameters (meaning that the three-parameter-fit $\chi^2$ value does not decrease significantly).
No significant change in the channel by channel size of the \brg\ line emitting region within the blue- and red-shifted \brg\ line components is observed. However, the values of the PA in the three-parameter fit do give some hint of different structures of the \brg\ line emitting region between the blue and red wings of the line. Presumably, the geometry is more complex than our simple models, and our results should be considered as an incentive to perform new observations with a better coverage of the $uv$ plane, allowing more complex models to be tested. 

Notably, when averaging the visibilities of the red- and blue-shifted line components over three spectral channels (Fig.\,\ref{fig:pure_line_visibilities_blue_red}.1 in Appendix\,\ref{appendixb:sec}), the averaged pure \brg\ line visibilities show indeed a difference between the blue- and red-shifted components for the three longest baselines. The corresponding averaged $a_{Br\gamma}$ over three spectral channels, at roughly $\sim$-34\,\kms\ and $\sim$75\,\kms on average, shows that the size of the blue-shifted \brg\ line emitting region ($a_{Br\gamma}$=0.58$\pm$0.07\,mas) is larger than the red-shifted one ($a_{Br\gamma}$=0.46$\pm$0.03\,mas). At a distance of 130\,pc, these values correspond to radii of 0.075$\pm$0.008\,au and 0.060$\pm$0.004\,au, which is much smaller than the inner rim size and the dust sublimation radius that is larger than 0.11\,au.

It should be noted that in addition to the statistical and observational errors, the absolute values of the continuum compensated \brg\ line visibilities could be also affected by the uncertainty on the assumed stellar parameters (e.g. spectral type, continuum flux). 
Taking this into account, we have recomputed the pure \brg\ line visibilities assuming a 10\% weaker line flux in the wings and a 10\% larger flux of the continuum (see Table\,\ref{tab:pure_visibilities}). Both scenarios only marginally affect the pure \brg\ line visibilities, and our main conclusions are not affected. 
Finally, it should be noticed than the differential phase values, as well as their errors, were taken into account when estimating the pure continuum compensated \brg\ line visibilities.  
Summarising, the difference in size between the blue- and red-shifted \brg\ line components is mainly due to the asymmetry in shape of the total visibilities with respect to the line profile. Furthermore, a different value of the assumed stellar parameters would not change the slope of the pure \brg\ line visibilities, as they would affect each spectral channel in the same way, especially at this low spectral resolution.

\begin{table*}
\centering
\caption{Geometric Gaussian fitting: pure \brg\ line visibilities.}
\label{tab:pure_line_visibilities}
\begin{tabular}{c|cc|ccccc}
\hline \hline
\multicolumn{1}{c}{}	& \multicolumn{2}{|c}{Single-parameter fit} & \multicolumn{4}{|c}{Three-parameter fit } \\
\multicolumn{1}{c}{RV}	& \multicolumn{1}{|c}{$\chi^2_r$} & \multicolumn{1}{c}{$a$\tablefootmark{a}}
& \multicolumn{1}{|c}{$\chi^2_r$} & \multicolumn{1}{c}{$a$\tablefootmark{a}} 
& \multicolumn{1}{c}{$i$} & \multicolumn{1}{c}{$PA$\tablefootmark{b}}  \\
$\mathrm{[km/s]}$ & & [mas] & & [mas] & [$^o$] & [$^o$] \\
\hline
-71 & 1.96	&   0.49 $\pm$  0.04 &  2.51  &  0.63 $\pm$ 0.16 & 32$^{+7}_{-9}$ & 162 $\pm$  19 \\
-35 & 2.03	&   0.51 $\pm$  0.03 &  2.09  &  0.66 $\pm$ 0.14 & 31$^{+7}_{-7}$ & 168 $\pm$  15 \\
3 & 1.45	&   0.54 $\pm$  0.03 &  2.54  &  0.54 $\pm$ 0.07 & 40$^{+4}_{-6}$ & 153 $\pm$  44 \\
39 & 0.60	&   0.48 $\pm$  0.03 &  0.92  &  0.47 $\pm$ 0.04 & 38$^{+7}_{-9}$ &  77 $\pm$  48 \\
75 & 2.12	&   0.46 $\pm$  0.04 &  2.92  &  0.46 $\pm$ 0.05 & 32$^{+12}_{-16}$ &  85 $\pm$  31 \\
112 & 2.38	&   0.47 $\pm$  0.06 &  3.39  &  0.47 $\pm$ 0.07 & 29$^{+12}_{-17}$ &  77 $\pm$  36 \\
\hline
\end{tabular}
\tablefoot{Single and three-parameter fit of pure Br$\gamma$ line visibilities at different radial velocities (${\rm RV}$).
\tablefoottext{a}{Size of the semimajor axis.}
\tablefoottext{b}{Position Angle of the semimajor axis with North up and East to the left. }
}
\end{table*}

\section{Discussion}
\label{sect:discussion}

\begin{figure}[t]
    \centering
    \includegraphics[width=7cm]{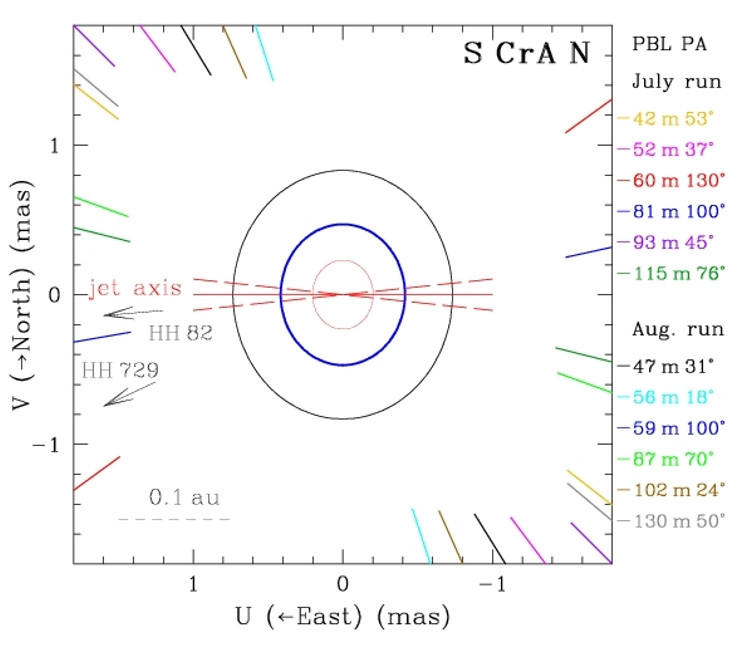}\\
    \includegraphics[width=6cm]{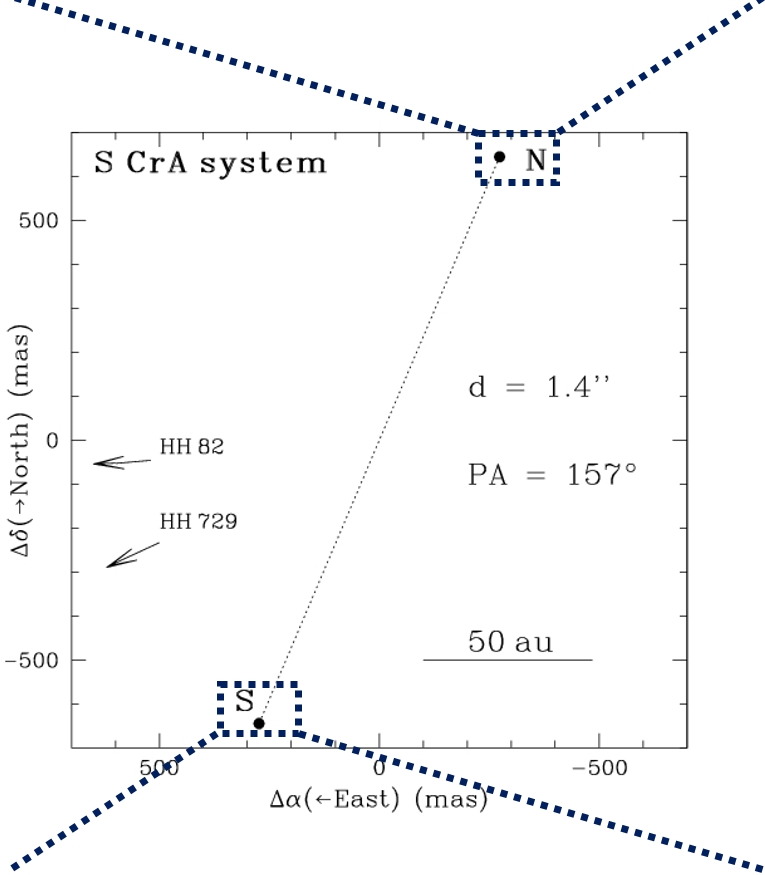}\\
    \includegraphics[width=7cm]{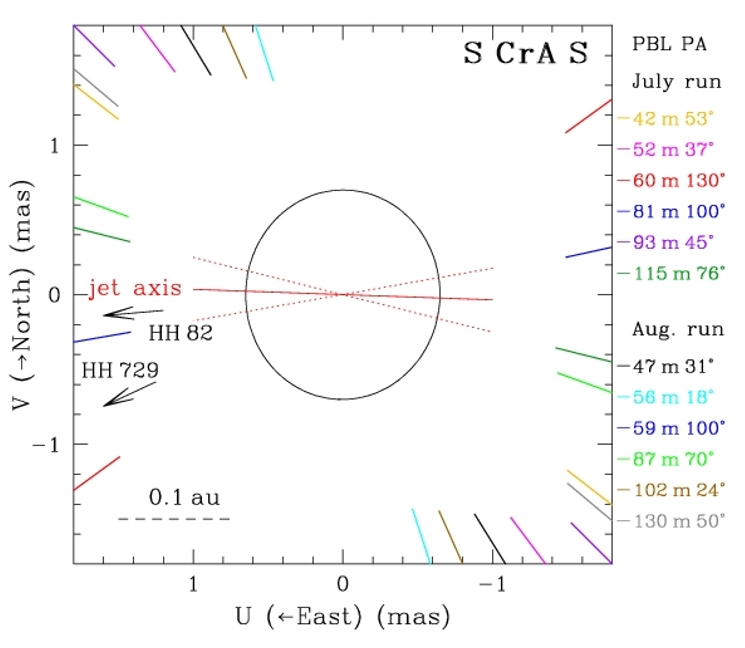}    \caption{{\bf Top:} Sketch of S\,CrA\,N as inferred from our GRAVITY observations. The gray and blue ellipses display the positions of the continuum and the Br$\gamma$ emitting regions, respectively. The red ellipse displays the inner truncation radius. The red dashed lines show the jet axis position with its uncertainty. Solid coloured lines show the orientation of the  baselines along with their lengths and position angles. {\bf Middle:} Sketch of the S\,CrA system. {\bf Bottom:} Sketch of S\,CrA\,S as inferred from our GRAVITY observations in the continuum.}
    \label{fig:sketch}
\end{figure}

Our K-band GRAVITY high spectral resolution results along with previous H- and K-band AMBER and PIONIER low resolution observations add to one of the best studied T Tauri stars in the near-infrared continuum dust and its innermost hydrogen circumstellar gas and represent, up to date, the first detailed interferometric study of the inner disk regions of a T Tauri binary system (see Fig.~\ref{fig:sketch}).

\subsection{Continuum emission}

The size derived for the continuum (r~=~0.108\,au) of S\,CrA\,N, is in excellent agreement with the values found by \cite{vural12} (r$\sim$0.11\,au) and \cite{PIONIER_TTauri} (r$\sim$0.108\,au). Moreover our estimates of the disk orientation and inclination ($PA$~=0\degr$\pm$6\degr and $i$~=28\degr$\pm$3\degr) are close to those proposed by \cite{Pontoppidan2011} for their disk wind model (i.\,e. $PA$~=15\degr and $i$~=10\degr, respectively). These values were obtained from the spectro-astrometric analysis of the CO fundamental lines at 4.7\,$\mu$m. For what concerns the spatial orientation of the binary system, the similarity in PA and $i$ of the disks of both components indicates that they were likely formed from the fragmentation of a common gravitationally unstable disk. Indeed, there are two main pathways to form binary or multiple systems: large scale fragmentation of turbulent cores and filaments or disk fragmentation \citep[see e.\,g.][]{adams89,bonnell94,fisher04}. There is direct and indirect observational evidence for both mechanisms \citep[see e.\,g.][]{Jensen2014,Tobin2016}. Direct observation mostly rely on ALMA imaging of the outer dusty disk (tens to hundreds of au), whereas indirect evidence mainly comes from the orientation of the jets (on scales from hundreds to thousands of au). Our analysis demonstrates that NIR interferometry can provide further evidence, giving direct information on the inner dusty disk geometry (down to sub-au scale) and on the system geometry of the binary system.

\subsection{Br$\gamma$ emission line in S\,CrA\,N}

Our GRAVITY-HR observations of S\,CrA\,N show a very compact, but resolved, \brg\ emitting region with a radius of $r \sim$0.06\,au, and thus, located well within the S\,CrA\,N dust sublimation radius which is larger than 0.11\,au. This small size places the origin of this line in the innermost regions of circumstellar disks, where different physical processes can contribute to its emission: magnetospheric accretion, wind, or heating of the hot upper layers of the disk. Similar results have also been found in other interferometric studies of Herbig Ae/Be and CTTSs \citep{kraus08, eisner14}. Indeed the most recent studies of Herbig Ae/Be stars have successfully modelled \brg\ interferometric observations using the disk surface \citep{ellerbroek14, mendigutia15}, a disk-wind, or a stellar-wind \citep{weigelt11,rebeca15,ale15,rebeca16,kurosawa16} as the main emitting mechanism for the \brg\ line. On the other hand, the origin of the \brg\ line in CTTSs is still uncertain \citep[][]{eisner10}. This is mostly due to the small size of the emitting region, which requires long baselines (of the order of 100\,m or more) to be spatially resolved, as well as high spectral resolution to analyse the different velocity components of the line.

Our GRAVITY observations add a fundamental ingredient to the interpretation of the origin of the \brg\ line and the physical processes taking place in the inner gaseous regions of CTTSs. For the first time, it has been possible to constraint the size of the emitting region and tentatively measure slight changes in the continuum-compensated pure \brg\ visibilities as a function of \textit{the line radial velocity} (Fig.\,\ref{fig:pure_line_visibilities_all_channels}). Our simple geometric Gaussian fitting points to slightly different dimensions of the region emitting the blue- and red-shifted \brg\ line components in S\,CrA\,N, with a red-shifted \brg\ emitting region slightly smaller than the blue-shifted one (r$_\mathrm{red}$~$\sim$0.060$\pm$0.004\,au vs. r$_\mathrm{blue}$~$\sim$0.075$\pm$0.008\,au). 
It should be noted, however, that our estimates of the continuum compensated \brg\ line visibilities are subject to unaccounted uncertainties such as the uncertainty on the spectral type of S\,CrA\,N  and the continuum level across the \brg\ line.
Despite this fact, the emitting region of the blue- and red-shifted \brg\ line components is, however, around a factor of two larger than the typical CTTS truncation radius of $r_\mathrm{trun}\sim$5\,R$_*$, that is $r_\mathrm{trun}\sim$0.03\,au assuming a stellar radius R$_*$ of 1.4\,R$_\odot$ \citep{petrov14_SCra}. This would suggest a wind origin of the \brg\ line in S\,CrA\,N, that would be supported by the presence of a large-scale protostellar jet arising from this source \citep{peterson11}. The base of a wind launched from the disk would initially rotate at the disk Keplerian rotation expected at the location of the footpoint of the wind. Assuming a stellar mass of M$_*\sim$1.5\,\msun\ \citep{prato03}, the disk Keplerian velocity at a distance from the star of 0.06\,au is ${\rm RV}_{\rm K}\sim$150\,\kms, which translates into a radial velocity of $\sim$70\,\kms assuming a disk inclination angle of $i=28^\circ$. This value is then in good agreement with both the \brg\ line FWHM ($\sim$180\,\kms) measured in S\,CrA\,N, and the velocity range (-71\,\kms<${\rm RV}$<112\,\kms) used to measure the size of the continuum-compensated \brg\ line emitting region. Notably, \cite{Pontoppidan2011} also model a disk-wind from their analysis of the CO fundamental lines in S\,CrA\,N. However, this wind is emitted at larger distance from the source ($\sim$100\,R$_*$ or $\sim$0.6\,au). Therefore it is reasonable to infer that we are observing the more compact atomic component of a disk-wind, that has a "onion-like" velocity structure, as proposed, for example, in DG\,Tau by \cite{vane14}.

On the other hand, given that the difference in size between the blue- and red-shifted \brg\ line components is real, our data exclude that the size asymmetry of the \brg\ line components might be due to the jet (orthogonal to the disk major axis PA, i.e. 90$\degr\pm$6$\degr$), as the visibility does not decrease for baselines roughly aligned with the jet PA. Moreover the analysed portion of the \brg\ spectrum should not trace the jet emission, likely traced by the high velocity component and possibly collimated at several au from the source~\citep[][]{ray07}, thus outside our FoV. We can also exclude that the small difference in size is due to a screening effect by optically thick foreground continuum, as the geometry of the system is almost face on. More likely, the slight size difference between the blue- and red-shifted \brg\ line components is due to a compact magnetosphere coexisting with the disk wind emission. As shown in previous interferometric studies, a magnetosphere would only slightly contribute to the total \brg\ line intensity, but it would modify the line profile along  the wings and increase the continuum-compensated \brg\ line visibilities of the red-shifted component \citep{rebeca15,rebeca16, kurosawa16,larisa16}. These characteristic features are, indeed, observed in S\,CrA\,N: namely the inverse P-Cygni profile in both \HeI\ and \brg\ lines, and a smaller size of the red-shifted \brg\ line emitting region. In contrast, while a disk wind can easily provide the shape of the blue wing, it cannot reproduce alone the red-shifted absorption component, which can be in turn easily shaped by magnetospheric accretion.
As already mentioned, previous optical observations of S\,CrA\,N display inverse P\,Cygni profiles in several \HI\ Balmer lines with radial velocities similar to that measured in the \brg\ line. In particular, \citet{bertout82} find that the red-shifted absorption feature is persistent during twelve consecutive nights. This supports the presence of an accretion column closely viewed pole-on~\citep{kurosawa13}, in agreement with the low inclination angle derived here.

\section{Conclusion}

Our GRAVITY observations of S\,CrA in the $K$-band continuum show two disk-like structures around each component of the binary system, with similar inclination and PA, suggesting that they originate from the fragmentation of a common parent disk. Both sources have comparable NIR continuum radii of $\sim$0.1\,au, while the dust sublimation radius is about 0.11-0.15\,au and 0.06-0.08\,au for the North and South components, respectively. Moreover, thanks to our observations at high spectral ($R$~$\sim$~4000) and spatial resolution, we managed to probe the innermost regions of the S\,CrA\,N component: its inner dusty disk encloses a compact \brg\ emitting region of r$\sim$0.06\,au, that is located in the inner gaseous disk but well beyond the truncation radius. Finally, our results indicate the presence of a wind and magnetospheric accretion traced by the \brg\ line in S\,CrA\,N, and for the first time, a tentative measure of slightly different dimensions of the region emitting the blue- and red-shifted components is attempted. Further observations of a sample of T~Tauri stars with different masses, ages and disk properties are planned with GRAVITY in the coming years with the aim of spatially resolving the hot (H~I) and warm (CO) gas in disks and investigating the structure, evolution and dynamics of disks of young stellar objects on sub-au scales.

\begin{acknowledgements}
These results are based on observations made with ESO Telescopes at the La Silla Paranal Observatory under programme IDs 60.A-9102. We thank the technical, administrative, and scientific staff of the participating institutes and the observatory for their extraordinary support during the development, installation, and commissioning of GRAVITY. R.G.L has received funding from the European Union’s Horizon 2020 research and innovation programme under the Marie Skłodowska-Curie Grant Agreement No. 706320. A.C.G. and T.P.R. were supported  by Science Foundation Ireland, grant 13/ERC/I2907. J.S.B acknowledges the support from the Alexander von Humboldt Foundation Fellowship Programme (Grant number ESP 1188300 HFST-P). K.P., B.L., M.B., and C.D. acknowledge the support of the French PNPS. This work has been supported by a grant from LabEx OSUG@2020 (Investissements d'avenir - ANR10LABX56). This research has made use of the Jean-Marie Mariotti Center \texttt{Aspro} and \texttt{SearchCal} services,  \footnote{Available at http://www.jmmc.fr/} and of CDS Astronomical Databases SIMBAD and VIZIER \footnote{Available at http://cdsweb.u-strasbg.fr/}.
\end{acknowledgements}

\bibliographystyle{aa} % style aa.bst
\bibliography{references.bib} % your references Yourfile.bib

%\Online

\begin{appendix}

\section{Interferometric observations of S\,CrA\,S around the \brg\ line and continuum visibilities}
\label{appendixa:sec}
%%%%%%%%%%%%%%%%%%%%%%%%%%%%%%%%%%%%%%%%%%%%%%%%%%%%%%%%%%%%%%%%%%%%%%%%%%%%%%%%%%%%%%%%%%%%%%%%%%%%%

\begin{figure*}[h]
	\includegraphics[width=\hsize]{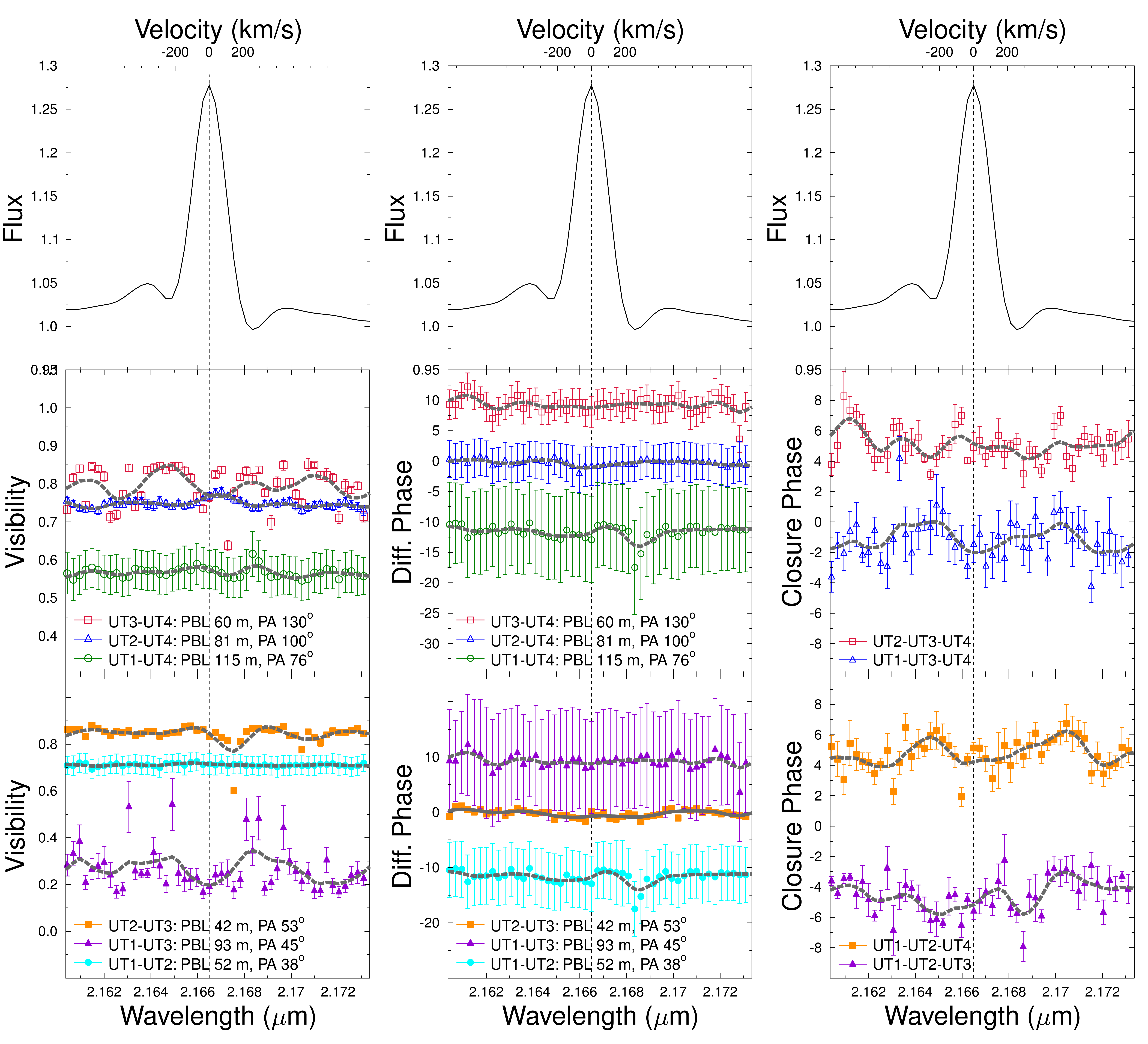}
    \caption{Same as Fig.\,\ref{fig:raw_data} but for S\,CrA\,S.}
    \label{fig:raw_dataB}
\end{figure*}

\section{Pure \brg\ line visibilities of S\,CrA\,N.}
\label{appendixb:sec}
\begin{figure*}[!th]
\centering
	\label{fig:pure_line_visibilities_blue_red}
	\includegraphics[width= 13cm, angle =0]{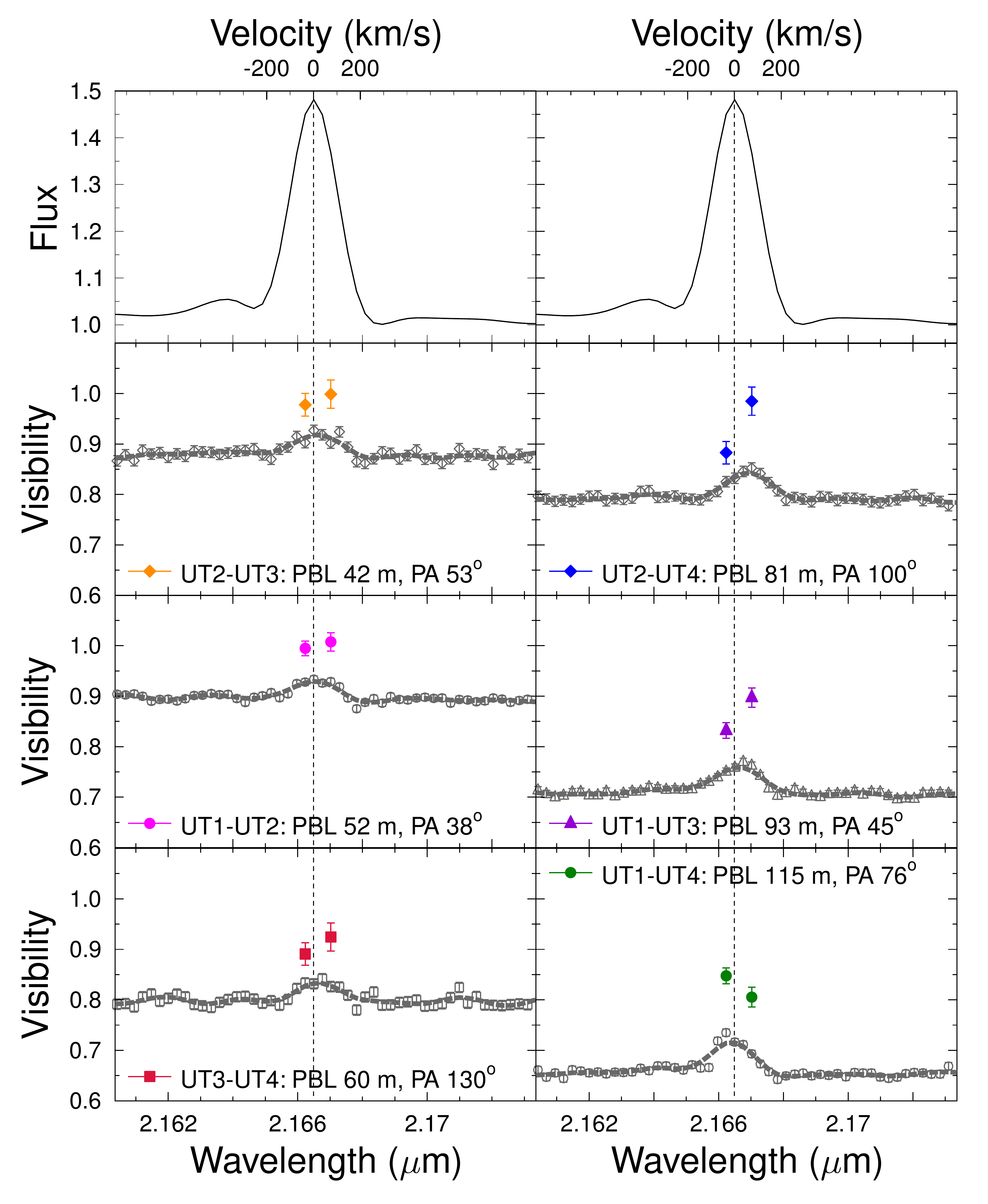}
	\caption{SCraA\,N continuum-compensated pure \brg\ visibilities (full coloured symbols) computed for each of
    the six GRAVITY baselines and averaged over three spectral channels. 
		}
\end{figure*}

\begin{figure*}[!th]
\centering
	\label{fig:fits_pure_line_visibilities}
	\includegraphics[width= \textwidth, angle =0]{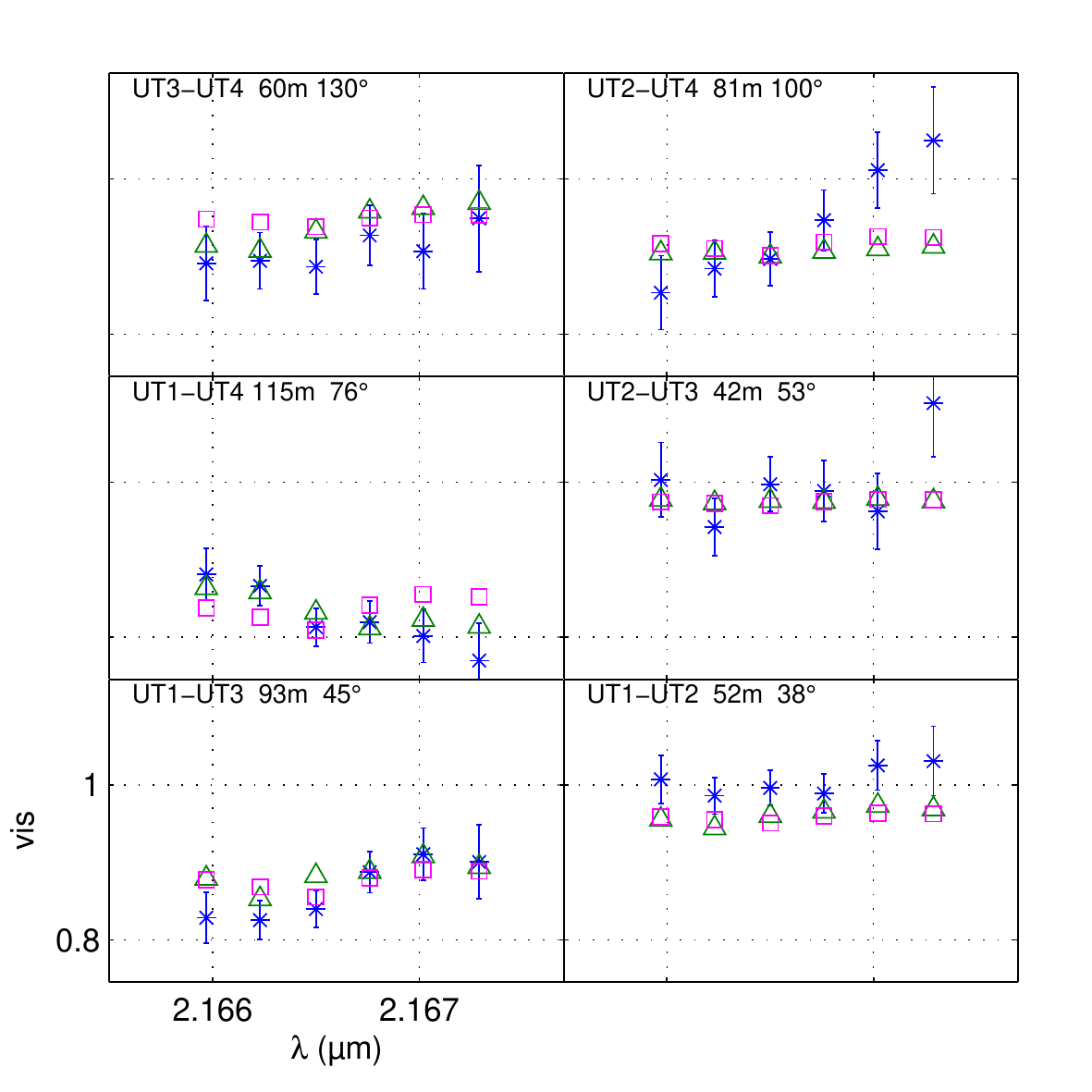}
	\caption{Best Gaussian fits of the S\,CrA\,N continuum-compensated pure \brg\ line visibilities (asterisks with error bars) computed for each of the six GRAVITY baselines. Open squares and triangles represent the one parameter (i.e. size) and three parameters fits (size, inclination and PA), respectively.}
\end{figure*}

\begin{table*}[h]
\caption{S\,CrA\,N continuum compensated pure \brg\ line visibilities.}
\label{tab:pure_visibilities}
\centering
%\vspace{0.2cm}
\begin{tabular}{c c c c | c c | c c}
\hline
\hline
PBL & PA & V$_{Br\gamma}^{blue}$\tablefootmark{a} & V$_{Br\gamma}^{red}$\tablefootmark{a} & V$_{Br\gamma}^{blue}$\tablefootmark{b} & V$_{Br\gamma}^{red}$\tablefootmark{b} & V$_{Br\gamma}^{blue}$\tablefootmark{c} & V$_{Br\gamma}^{red}$\tablefootmark{c} \\ 
 $[m]$         & [$\circ$]    &   & &  &  \\ 
\hline
42  & 53  & 0.98$\pm$0.02 & 1.00$\pm$0.02 & 1.00$\pm$0.03 & 1.00$\pm$0.03 & 1.00$\pm$0.03 & 1.00$\pm$0.03  \\
52  & 38  & 0.99$\pm$0.01 & 1.00$\pm$0.02  &1.00$\pm$0.02 & 1.00$\pm$0.02 & 1.00$\pm$0.02 & 1.00$\pm$0.02\\
60  & 130 & 0.89$\pm$0.02 & 0.92$\pm$0.03  & 0.99$\pm$0.02 & 1.00$\pm$0.03 & 0.91$\pm$0.03 & 0.96$\pm$0.03\\
81  & 100 & 0.88$\pm$0.02 & 0.98$\pm$0.03  & 0.98$\pm$0.02 & 1.00$\pm$0.03 & 0.90$\pm$0.02  & 1.00$\pm$0.04\\
93  & 45  & 0.83$\pm$0.01 & 0.90$\pm$0.02  & 0.93$\pm$0.02 & 1.00$\pm$0.02 & 0.86$\pm$0.02 & 0.95$\pm$0.02 \\
115 & 76  & 0.85$\pm$0.01 & 0.80$\pm$0.02  &0.94$\pm$0.02 & 0.89$\pm$0.02 &  0.89$\pm$0.02  &  0.85$\pm$0.02 \\
\hline
\end{tabular}
\tablefoot{
\tablefoottext{a}{Averaged continuum compensated pure \brg\ visibilities over three spectral channels at $\sim$-34\,\kms\ (V$_{Br\gamma}^{blue}$) and $\sim$+75\,\kms\ (V$_{Br\gamma}^{red}$).}
\tablefoottext{b}{As \emph{a} but computed assuming a decrease of the line flux of 10\% with respect to the measured one.}
\tablefoottext{c}{As \emph{a} but computed assuming an increase of the continuum flux of 10\% with respect to the measured one.}
}
\end{table*}

\end{appendix}

\end{document}